\documentclass{article}

\usepackage{PRIMEarxiv}

\usepackage[utf8]{inputenc} % allow utf-8 input
\usepackage[T1]{fontenc}    % use 8-bit T1 fonts
\usepackage{hyperref}       % hyperlinks
\usepackage{url}            % simple URL typesetting
\usepackage{booktabs}       % professional-quality tables
\usepackage{amsfonts}       % blackboard math symbols
\usepackage{nicefrac}       % compact symbols for 1/2, etc.
\usepackage{microtype}      % microtypography
\usepackage{lipsum}
\usepackage{xspace}
\newcommand{\ie}{\textit{i.e.}}
\newcommand{\eg}{\textit{e.g.}}
\newcommand{\apriori}{\textit{a priori}}
\usepackage{fancyhdr}       % header
\usepackage{graphicx}       % graphics
\graphicspath{{media/}}     % organize your images and other figures under media/ folder
\usepackage[super,numbers]{natbib}

\usepackage[inline]{enumitem}
\usepackage[ruled,vlined]{algorithm2e}
\SetAlCapNameFnt{\small}
\SetAlCapFnt{\small}
\usepackage{algpseudocode}

\usepackage{amssymb}
\usepackage{float}
\usepackage{amsmath}
\usepackage{chemarrow}
\usepackage{array}
\usepackage{placeins}
\usepackage{nicefrac}

\usepackage{xfrac}
\usepackage{siunitx}
\usepackage[version=4]{mhchem}

\usepackage{longtable}
\usepackage{multirow}
\usepackage{array}
\usepackage{boldline}
\usepackage[flushleft]{threeparttable}
\usepackage{booktabs}

\title{Deep Kernel Bayesian Optimisation for Closed-Loop Electrode Microstructure Design with User-Defined Properties based on GANs\thanks{
\textit{Authors' emails:} \textbf{A. Gayon-Lombardo}: \texttt{andrea.gl.6392@gmail.com}, \textbf{E.A. del Rio-Chanona}: \texttt{a.del-rio-chanona@imperial.ac.uk}, \textbf{C.A. Pino-Mu\~{n}oz}: \texttt{c.pino15@imperial.ac.uk}, \textbf{N.P. Brandon}: \texttt{n.brandon@imperial.ac.uk}\\
ORCIDs: 0000-0001-9907-3565 (A. Gayon-Lombardo), 0000-0003-0274-2852 (E.A. del Rio-Chanona), 0000-0001-5138-4030 (C.A. Pino-Mu\~{n}oz), 0000-0003-2230-8666 (N.P. Brandon)
}
}

\author{
  Andrea Gayon-Lombardo \\
  Department of Earth Science and Engineering \\
  Imperial College London\\
  SW7 2AZ, United Kingdom\thanks{\textit{Author's current affiliation}:AFRY MCD, email: \texttt{andrea.gayonlombardo@afry.com} }
   \And
  Ehecatl A. del Rio-Chanona \\
  Department of Chemical Engineering \\
  Imperial College London \\
  SW7 2AZ, United Kingdom
  \AND
  Catalina A. Pino-Mu\~{n}oz \\
  Department of Earth Science and Engineering \\
  Imperial College London \\
  SW7 2AZ, United Kingdom
  \And
  Nigel P. Brandon \\
  Department of Earth Science and Engineering \\
  Imperial College London \\
  SW7 2AZ, United Kingdom
}

\begin{document}
\maketitle

\begin{abstract}
The generation of multiphase porous electrode microstructures with optimum morphological and transport properties is essential in the design of improved electrochemical energy storage devices, such as lithium-ion batteries. Electrode characteristics directly influence battery performance by acting as the main sites where the electrochemical reactions coupled with transport processes occur. This work presents a generation-optimisation closed-loop algorithm for the design of microstructures with tailored properties. A deep convolutional Generative Adversarial Network is used as a deep kernel and employed to generate synthetic three-phase three-dimensional images of a porous lithium-ion battery cathode material. A Gaussian Process Regression uses the latent space of the generator and serves as a surrogate model to correlate the morphological and transport properties of the synthetic microstructures. This surrogate model is integrated into a deep kernel Bayesian optimisation framework, which optimises cathode properties as a function of the latent space of the generator. A set of objective functions were defined to perform the maximisation of morphological properties (\eg, volume fraction, specific surface area) and transport properties (relative diffusivity). We demonstrate the ability to perform simultaneous maximisation of correlated properties (specific surface area and relative diffusivity), as well as constrained optimisation of these properties. This is the maximisation of morphological or transport properties constrained by constant values of the volume fraction of the phase of interest. Visualising the optimised latent space reveals its correlation with morphological properties, enabling the fast generation of visually realistic microstructures with customised properties.
\end{abstract}

% keywords can be removed

\section{Introduction}
\label{sec:intro} 

Electrochemical energy storage systems (EESS) are essential for decarbonising the electricity grid. They provide grid-balancing capabilities that enable the integration of intermittent renewable energy sources. Therefore, advancing their efficiency and scalability is pivotal for achieving global carbon neutrality goals. Electrodes constitute the core component of EESS since they are the sites where electrochemical reactions occur. To fully harness the potential of EESS, a focus on the development and improvement of electrodes and the optimisation of their microstructure is required. \cite{Moussaoui2018,Lu2020}.

These electrodes are composed of multiple phases, including solid materials that conduct charge, and void space (\ie, porosity) that allow a maximum active surface area, but need to contain percolating paths to enable ionic and electronic transport while maintaining sufficient mechanical integrity~\cite{Moussaoui2019, Cooper2017}. Enhancing the electronic and ionic transport within electrodes' conducting pathways can lead to high utilisation of the active materials \cite{Bielefeld2019}. This results in electrodes with higher performance. A thorough understanding of the interplay between electrodes' transport processes and microstructural properties is crucial for designing next-generation EESS.

The effect of electrode morphology and microstructural properties on performance has been analysed by micro-scale level simulation of the physical and electrochemical processes~\cite{Lu2020,Boyce2022}. For example, high porosity and low tortuosity electrodes have been evidenced to impact the ionic conductivity by enhancing the transport of Lithium ions through the liquid electrolyte in Li-ion batteries~\cite{Li2021}. Nonetheless, increased porosity would lead to a reduced volume fraction of the active material and, thus, a reduced battery energy density~\cite{Lu2020}. 
This suggests that the interplay between microstructural properties must be considered when optimising the electrode's morphology. Apart from micro-scale models, continuum modelling has been implemented to elucidate the optimum porosity, the effect of graded porosity, and the optimum effective diffusivity~\cite{Suthar2015,Lu2020}. However, these optimum values are theoretical and result in models where the microstructure obtained is idealised.

Although these modelling efforts are insightful, a step involving the quantification of the electrode spatial arrangement or geometry in correlation to its estimated microstructural properties (\ie, porosity, specific surface area, tortuosity) is critical to understanding the microstructure-performance relationship as a closed-loop process~\cite{Zhang2019}. 
The key question lies in the existence of a mathematical function that can define the electrode spatial arrangement, which can be manipulated based on a set of parameters to obtain ideal or user-specified properties that can maximise the performance. One pathway for analysing the microstructure-performance relation consists of generating synthetic electrodes through statistical and computational modelling. Generating synthetic multiphase electrode microstructures could provide insight into the optimum morphology required for designing high-performance electrodes.

A large body of work has been presented in generating synthetic microstructure for energy materials. A statistical method consisting of two-point correlation function was introduced initially by Suzue et al.~\cite{Suzue2008} and implemented by Baniassad et al.~\cite{Baniassadi2011} to reconstruct three-dimensional microstructures of composite Solid Oxide Fuel Cell (SOFC) anodes. Alternative algorithms introduced by Ali et al.~\cite{Ali2008} and Kenney et al.~\cite{Kenney2009} implement a sphere packing and growth technique to simulate the sintering process for synthesising SOFC electrodes. This method was later implemented by Bertei et al. ~\cite{Bertei2012} and Cai et al.~\cite{Cai2011} to reconstruct a 3D micorstructure of  SOFC anode and model its percolating behaviour and electrochemical performance respectively. In Li-ion batteries, previous authors have performed particle-based simulations to correlate the fabrication process of electrodes to their respective microstructure. Forouzan et al.~\cite{Forouzan2016} modeled the synthesis process of a Li-ion cathode to predict its microstructural and mechanical properties. Srivastava et al.~\cite{Srivastava2020} implemented a particle-based simulation to analyse the effect of carbon binder adhesion in the electrode microstructural and transport properties. 

These physics-based models can predict the effect of the microstructure on transport properties; however, they have proved to be computationally expensive and specific to a particular type of electrode material.Recent advances in deep learning have led to the implementation of generative models for the stochastic generation of porous media. \cite{Mosser2017} implemented a Generative Adversarial Network (GAN) to reconstruct the three-dimensional microstructure of two-phase synthetic and granular materials. \cite{Gayon-Lombardo2019} extended this method to generate three-dimensional, three-phase electrodes: a Solid-Oxide Fuel Cell anode and a Li-ion cathode. These reconstructions possessed the same microstructural properties and two-point correlation function as the original tomographic data.

In comparing the most common microstructure generation techniques, GANs can perform fast sampling of high-dimensional and intractable density functions without the need for an \apriori~model of the probability distribution function to be specified~\cite{Mosser2018}. Thus, GANs proved to be an efficient method for generating realistic microstructures, where the trained generator constitutes a virtual representation of the real microstructure.

The GAN generator takes a latent vector ${\bf z}$ as its input and maps it to the target data space in order to produce diverse outputs~\cite{Gayon-Lombardo2020}. These latent vectors are sampled from a standard distribution, such as a normal distribution. The capabilities of the latent space ${\bf z}$ of GANs and its correlation with the output image have been explored~\cite{Bojanowski2019,Li2018}. By interpolating between pairs of ${\bf z}$ vectors, the generator is able to produce semantically meaningful images and a smooth transition between each other~\cite{Bojanowski2019}. Moreover, linear arithmetic operations can be performed in the latent space of GANs which leads to meaningful transformations of images with visually different properties~\cite{Bojanowski2019}. It is therefore clear that the output image from the trained generator will be directly correlated to the input vector in the latent space. Any new microstructure obtained from a trained generator using an input ${\bf z}$ from the latent space distribution
would be visually different but will possess the same distribution of statistical and microstructural properties as the training set~\cite{Mosser2018,Mosser2017}. If the latent vector input deviates from a normal distribution, the generated microstructure may have different properties (\eg, volume fractions, diffusivities) than the training set but remains visually realistic and indistinguishable from real data~\cite{Gayon-Lombardo2020}. Based on this, an optimisation of the latent space can be performed in order to obtain a microstructure with a set of desired properties~\cite{Li2018}. 

The electrode's microstructural and transport properties are conventionally calculated using physics-based simulations. However, these simulations are computationally expensive and cannot be back-propagated to the latent variables in a gradient-based optimisation. Thus, a surrogate model must be defined that correlates the input latent vector $({\bf z})$ with the properties of the generated microstructure ${f(\bf z})$.

Recent interest has risen to implement Gaussian Processes (GP) as surrogate models to represent these complex microstructures since they provide a powerful tool for derivative-free optimisation \cite{Li2018,Yang2018}. GP regressions were first proposed by \cite{Ohagan1978} and then popularised by \cite{Neal2012} and \cite{Rasmussen2006}. GPs are non-parametric models used to make predictions about complex systems with uncertainty, provided enough data is available~\cite{Bradford2020}. \cite{Jung2020} proposed a 3D convolutional autoencoder for the generation of a two-phase steel microstructure and optimised its latent space using Bayesian optimisation with GP to achieve maximum tensile strength. Similarly, \cite{Yang2018} proposed a workflow using Bayesian optimisation with GP for optimising the microstructure generated using the generator of a trained GAN. They demonstrated the applicability of this methodology for optimising the optical properties of a two-phase material. Such deep kernel Bayesian optimisation (DK-BO) combines deep kernel learning, which merges the non-parametric flexibility of kernel methods with the structural properties of deep neural networks, offering a more expressive alternative to standard GPs~\cite{Wilson2015,Bowden2021}. This approach helps address the key limitation of Bayesian optimisation of scaling to higher-dimensional problems where the number of required function evaluations increases exponentially. In a typical Bayesian optimisation framework, a GP surrogate is trained and the next sampling point is selected by optimising an acquisition function. Compared to traditional GPs, DK learning models are more expressive and can learn better representations.

In this paper, we introduced a deep kernel Bayesian optimisation framework that enables a closed-loop generation-optimisation process, integrating synthetic microstructure generation with the optimisation of user-specified properties for a three-phase, three-dimensional porous microstructure. Our goal is to design a multiphase electrode with optimum user-specified microstructural and transport properties based on open-source tomographic data. As a case study, we employ the GAN generator previously developed by the authors~\cite{Gayon-Lombardo2019} to generate optimised cathode microstructures for Lithium-ion batteries. This work first implements an unconstrained Bayesian optimisation of the input latent vector $\bf z$, while a physics-based model evaluates the microstructural and transport properties of the optimised cathodes. Here, the GP regression method is used as the surrogate model that maps the inputs ${\bf z}$ and outputs ${f(\bf z})$. Then, we explore the trade-offs between the microstructural and transport properties to propose and test constrained optimisation strategies that can lead to better performing cathodes. 

%%% METHODS %%%
\section{Methods}
\label{sec:Method}
\subsection{Generative Adversarial Network}

GANs are deep generative models which can learn the probability distribution functions of a given data set by training two functions \--- a generator, $G_{\theta}(\mathrm{\textbf{z}})$, and a discriminator, $D_{\theta}(\mathrm{\textbf{x}})$ \citep{Goodfellow2014}. A detailed explanation of the GAN methodology can be found in the authors' previous work~\citep{Gayon-Lombardo2020}; in this section, we only include the general idea.

Both functions, generator and discriminatory, were taken to be deep convolutional neural networks. The discriminator's loss function, $J^{(D)}$, is defined by

\begin{equation}
J^{(D)} = J^{(D)}_\mathrm{BCE} + J^{(D)}_\mathrm{MSE}, \label{eq:JD}
\end{equation}

\noindent and comprises two terms. One that corresponds to the binary cross-entropy (BCE) loss in a classifier between the discriminator's prediction and the real label,

\begin{equation}
    \begin{aligned}
    J^{(D)}_\mathrm{BCE} =& \mathbb{E}_{\mathrm{\textbf{x}}\sim p_{\mathrm{data}}(\mathrm{\textbf{x}})} \left[\log \left(D_\theta(\mathrm{\textbf{x}})\right) \right] \\
    &+ \mathbb{E}_{\mathrm{\textbf{z}}\sim p_{\mathrm{z}}(\mathrm{\textbf{z}})} \left[\log \left(1 - D_\theta\left(G_\theta(\mathrm{\textbf{z}})\right) \right) \right],\label{eq:JBCE}
    \end{aligned}
\end{equation}

\noindent and a second term corresponding to the reconstruction loss between the real data ${\bf x}$ and the generated data $G_\theta({\bf z})$ in order to increase the resolution of the synthetic realisations, 

\begin{equation}
J^{(D)}_\mathrm{MSE} = \frac{1}{N} \sum_{i=1}^N \left({\bf x} - G_\theta({\bf z})\right)^2.\label{eq:JMSE}
\end{equation}

\noindent where, ${\bf z}$ is the latent space. The subscript BCE refers to binary cross-entropy error, and MSE refers to mean squared error.

The generator was trained to minimise a loss function, $J^{(G)}$, comprising the log-probability of the discriminator being correct, 

\begin{equation}
J^{(G)} = \mathbb{E}_{\mathrm{\textbf{z}}\sim p_{\mathrm{z}}(\mathrm{\textbf{z}})} \left[\log \left(1 - D_\theta\left(G_\theta(\mathrm{\textbf{z}})\right) \right) \right].\label{eq:JG}
\end{equation}

An optimisation problem defined by a Nash equilibrium was performed until each player (\ie, discriminator and generator) achieved a local minimum and the discriminator could not distinguish between real and synthetic data \citep{Goodfellow2016}.

\subsection{Gaussian Process regression}
Gaussian Processes (GPs) are generalisations of a multivariate Gaussian distribution to infinite dimensions. A GP regression aims to model an unknown latent function, $f(\mathrm{\textbf{x}})$, using noisy observations, $y$, when an arbitrary input vector $\mathrm{\textbf{x}} \in \mathbb{R}^n$ is considered~\citep{Ebden2008,Jones1998},

\begin{equation}
    y = f(\mathrm{\textbf{x}}) + \epsilon, \; \;   \epsilon \sim \mathcal{N}\left(0,\sigma^2_\epsilon\right).
    \label{eq:y_distribution}
\end{equation}

\noindent A measurement of the noise of the Gaussian distribution is the error, $\epsilon \in \mathbb{R}$, which has zero mean and variance $\sigma^2_\epsilon$~\citep{Rasmussen2006,Rasmussen1997}. 

We assume that $f({\bf x})$ follows a Gaussian process and therefore can be modelled as 

\begin{equation}
 f({\bf x}) \sim  GP\left(m(\mathrm{\textbf{x}}), k(\mathrm{\textbf{x}},\mathrm{\textbf{x}}') \right).\label{eq:fx}\\
\end{equation}

\noindent A mean function, $m(\mathrm{\textbf{x}})$, and a covariance function, $k(\mathrm{\textbf{x}},\mathrm{\textbf{x}}')$ fully specified this GP,

\begin{align}
m(\mathrm{\textbf{x}}) &= \mathbb{E}\left[f(\mathrm{\textbf{x}})\right],\label{eq:mean}\\
k(\mathrm{\textbf{x}},\mathrm{\textbf{x}}') &= \mathbb{E}\left[(f(\mathrm{\textbf{x}}-m(\mathrm{\textbf{x}}))f(\mathrm{\textbf{x}}'-m(\mathrm{\textbf{x}}'))^T\right].\label{eq:covariance}
\end{align}

A non-parametric regression model can be implemented assuming that the function $f(\mathrm{\textbf{x}})$ is a sample from a GP~\citep{Richardson2017,Richardson2019}. The noisy observations, $y$, must also follow a GP with the same mean, $m(\mathrm{\textbf{x}})$, but with a different covariance function to account for the measurement of noise,  

\begin{equation}
    \begin{aligned}[t]
    &y \sim GP\left(m(\mathrm{\textbf{x}}), k(\mathrm{\textbf{x}},\mathrm{\textbf{x}}') + \sigma^2_\epsilon \delta(\mathrm{\textbf{x}},\mathrm{\textbf{x}}') \right),\label{eq:GP}\\
    &\textrm{where, }\delta(\mathrm{\textbf{x}},\mathrm{\textbf{x}}') = \begin{cases} 
    1 & \text{if } \mathrm{\textbf{x}} = \mathrm{\textbf{x}}'\\
    0 & \text{if } \mathrm{\textbf{x}} \neq \mathrm{\textbf{x}}'\end{cases},
\end{aligned}
\end{equation}

\noindent as expected by the additive property of Gaussian distributions. Here, $\delta$ is the Kronecker-delta. 

The prior of the function, which would be updated based on input-output data observations, is defined by Eqs.~\ref{eq:fx} and~\ref{eq:GP}. Both a zero-mean, commonly used in Machine Learning~\cite{Bradford2018,Bradford2020,Bradford2018b}, and a squared-exponential (SE) covariance function, which is the stationary covariance frequently applied, were used~\cite{Rasmussen2006,Ohagan1978}, 

\begin{equation}
    k_{SE}(x_i,x_j)=\sigma_f^2~\textrm{exp} \left(-\frac{1}{2\lambda_d^2}(x_i-x_j)^2 \right).
    \label{eq:SE}
\end{equation}

\noindent The covariance function hyper-parameters, $\sigma_f^2$ and $\lambda_d^2$, control the $y$-scaling and $x$-scaling of the function, respectively. For multiple dimensions, Eq.~\ref{eq:SE} can be rewritten as Eq.~\ref{eq:multi_SE}, where $W = \mathrm{diag}[w_1, \dots, w_D]$, with elements $w_d = 1/\lambda_d^2$ is used.

\begin{equation}
    k_{SE}(\mathrm{\textbf{x}}_i,\mathrm{\textbf{x}}_j)=\sigma_f^2~\textrm{exp} \left(-\frac{1}{2}(\mathrm{\textbf{x}}_i-\mathrm{\textbf{x}}_j)^T W(\mathrm{\textbf{x}}_i-\mathrm{\textbf{x}}_j) \right)    
    \label{eq:multi_SE}
\end{equation}

To explain the GP implementation, $N$ available observations $\mathrm{\textbf{y}} = \left[y_1, \dots, y_N\right]^T$ for $N$ different inputs $\mathrm{\textbf{X}} = \left[\mathrm{\textbf{x}}_1, \dots, \mathrm{\textbf{x}}_N\right]$ were considered. Each element in ${\bf y}$ is a scalar value, and ${\bf y}$ corresponds to all $N$ values (or samples) concatenated. The uncertainty of estimating $N$ function values can be represented based on the prior from the mean and the covariance function of the vector $\mathrm{\textbf{F}} = \left[f(\mathrm{\textbf{x}}_1), \dots, f(\mathrm{\textbf{x}}_N)\right]^T$. This vector has a mean vector defined as \textbf{0} and a $N \times N$ covariance matrix, $\Sigma$,

\begin{equation}
    \Sigma := \Sigma(\textbf{x}_i,\textbf{x}_j) 
    \label{eq:covF}
\end{equation}

\noindent The uncertainty of the observation matrix $\mathrm{\textbf{y}}$ can be expressed in the same way as $\mathrm{\textbf{F}}$ with a mean function of \textbf{0} and a covariance matrix, $\textbf{K}$, defined by, 

\begin{equation}\label{eq:total_cov}
    \textbf{K} =
    \left[\Sigma(\textbf{x}_i,\textbf{x}_j)  + \sigma_\epsilon^2\delta(\mathrm{\textbf{x}}_i,\mathrm{\textbf{x}}_j) \right]_{N \times N}
    = \Sigma + \sigma_\epsilon^2\textbf{I}.    
\end{equation}

\noindent Here, $\sigma_\epsilon^2$ is the variance of Gaussian distributed perturbation noise as given by Eq.~\ref{eq:y_distribution}.

Considering the training data $\mathrm{\textbf{X}}$ and $\mathrm{\textbf{y}}$, a GP is fully defined by the hyper-parameters of the covariance function and the random noise of $y$ in Eq.~\ref{eq:total_cov}. These hyper-parameters are commonly unknown \textit{a priori}, and therefore, a step for estimating their value must be considered in the Gaussian Process regression. They are jointly denoted by the vector ${\bf\Theta} = [w_1, \dots, w_D, \sigma_f^2, \sigma_\epsilon^2]$, and can be estimated by maximising the log-likelihood of the conditional probability density function (PDF) $p\left(\mathrm{\textbf{y}}|\mathrm{\textbf{X}},{\bf\Theta}\right)$. It is known that, 
$\mathrm{\textbf{y}}|\mathrm{\textbf{X}}, {\bf\Theta} \sim \mathcal{N}(\textbf{0},\mathrm{\textbf{K}})$, and therefore the PDF can be defined by, 

\begin{equation}
    p({\bf y| X, \Theta})=\frac{1}{(2\pi)^{n/2}(\mathrm{det}[{\bf K}])^{1/2}}\mathrm{exp}\left[-\frac{1}{2}{\bf Y}^T{\bf K}^{-1}{\bf y}\right].
    \label{eq:conditional_prob}
\end{equation}

\noindent The log-likelihood of the PDF was obtained by applying the log function,

\begin{equation}
    \textrm{log}[p({\bf y| X,\Theta})]=-\frac{n}{2}\textrm{log}(2\pi)-\frac{1}{2}\textrm{log}(\textrm{det}({\bf K}))-\frac{1}{2}{\bf y}^T{\bf K}^{-1}{\bf y}.
    \label{eq:log_pdf}
\end{equation}

\noindent Although hyper-parameters, ${\bf \Theta}$, are not explicitly shown in Eq.~\ref{eq:log_pdf}, they are contained in $\mathrm{\textbf{K}}$ (according to Eq.~\ref{eq:total_cov}). A convenient way of maximising the log-likelihood is to minimise the negative log-likelihood (NLL) \cite{Rasmussen2006,Sundararajan2000}, thus, the negative of Eq.~\ref{eq:log_pdf} was minimised,

\begin{equation}
    \mathcal{L}({\bf \Theta})=-\textrm{log}[p({\bf y| X,\Theta})].
    \label{eq:NLL}
\end{equation}

\noindent A non-linear optimisation algorithm was implemented to obtain the optimum values of the hyper-parameters, ${\bf \Theta}^*$ based on the training data,

\begin{equation}
    {\bf \Theta}^* \in \textrm{argmin}_{\bf \Theta}~\mathcal{L}({\bf \Theta}).
    \label{eq:NLL_opt}
\end{equation}

\noindent Then, predictions were inferred at unknown input data, $\mathrm{\textbf{x}}_{new}$, by computing the conditional probability distribution of the Gaussian process on data ${\bf X}$,${\bf y}$,${\bf x}_\mathrm{new}$. For the covariance of observations ${\bf y}$ given by Eq.~\ref{eq:total_cov}, a mean function of $m({\bf x}) = 0$ was chosen. When a new observation ${\bf x}_\mathrm{new}$ was taken, the prediction $y_\mathrm{new}$ was conditioned to the observed training data,

\begin{equation}
    \begin{bmatrix}  {\bf y}  \\  y_\mathrm{new} \end{bmatrix}
\sim\mathcal{N}
\begin{pmatrix} \begin{bmatrix}  {\bf 0}  \\  0 \end{bmatrix},\begin{bmatrix}  {\bf K} & {\bf \Sigma}({\bf X},{\bf x}_\mathrm{new})  \\  {\bf \Sigma}({\bf x}_\mathrm{new},{\bf X}) & {\bf \Sigma}({\bf x}_\mathrm{new},{\bf x}_\mathrm{new}) \end{bmatrix}\end{pmatrix},
\label{eq:cond_prob}
\end{equation}

\noindent where,
\begin{equation}
    {\bf \Sigma}({\bf X},{\bf x_\mathrm{new}})=[k({\bf x}_1,{\bf x}_\mathrm{new}),...,k({\bf x}_n,{\bf x}_\mathrm{new})]^T.
\end{equation}

The conditional probability of the prediction for the new observation $p(y_\mathrm{new}|{\bf X},{\bf y},{\bf x}_\mathrm{new})$ was obtained analytically by applying the conditional rules of Gaussian distributions given by, 

\begin{equation}
    y_\mathrm{new}|{\bf X},{\bf y},{\bf x}_\mathrm{new}\sim\mathcal{N}(\mu_{y_\mathrm{new}},\sigma^2_{y_\mathrm{new}}),
    \label{eq:y_new}
\end{equation}

\noindent where, $\mu_{y_\mathrm{new}}$ is the prediction and $\sigma_{y_\mathrm{new}}$ is the variance of the corresponding uncertainty,

\begin{equation}
\begin{aligned}
\mu_{y_\mathrm{new}} = & \mathbb{E}[y_\mathrm{new}|{\bf X},{\bf y},{\bf x}_\mathrm{new}] \\
= & {\bf \Sigma}({\bf x}_\mathrm{new},{\bf X})[{\bf \Sigma}({\bf X},{\bf X}) + \sigma_\epsilon^2I]^{-1}{\bf y},
\end{aligned}
\label{eq:mean_new}  
\end{equation}

\noindent and,

\begin{equation}
\begin{aligned}
    \sigma_{y_\mathrm{new}}= & {\bf \Sigma}({\bf x}_\mathrm{new},{\bf x}_\mathrm{new})\\
    &-{\bf \Sigma}({\bf x}_\mathrm{new},{\bf X})[{\bf \Sigma}({\bf X},{\bf X}) + \sigma_\epsilon^2I]^{-1}{\bf \Sigma}({\bf X},{\bf x}_\mathrm{new}).
\end{aligned}
\label{eq:cov_new}
\end{equation}

This GP implementation mapped the correlation between the training data set of inputs-outputs, and in order to obtain predictions at arbitrary inputs, a three-step procedure was followed:
\begin{enumerate}
    \item mean and covariance function were selected based on the prior knowledge of the function to model. In some cases, an explicit mean function can be implemented in order to capture the prior information of the expected form of the model~\citep{Ebden2008,Richardson2017,Richardson2019}. A zero mean was implemented in this work;
    \item hyper-parameters were optimised by minimizing the negative log-likelihood using the training set~\citep{Sundararajan2000}; and
    \item prediction of a new observation for an arbitrary input was obtained using Eqs.~\ref{eq:mean_new} and~\ref{eq:cov_new}~\citep{Bradford2018}.
\end{enumerate}

\subsection{Bayesian Optimisation with Gaussian Processes}

Gaussian Processes are effective to model an objective function while taking uncertainty explicitly into account, and therefore they are ideal methods for expensive black-box optimisations.\cite{Jones1998} The GP regression not only provides an accurate prediction of unknown outputs but also presents a measure for predicting uncertainty. This poses a significant advantage compared to commonly used black-box optimisation methods and makes the GP regression a powerful tool for derivative-free optimisation, both for single-objective optimisation \cite{Jones1998,Shahriari2016} and multi-objective optimisation \cite{Bradford2018,Bradford2020,Bradford2018b}. The optimisation problem to find the values of ${\bf x}$ that minimise the function $f({\bf x})$ was defined as

\begin{equation}
\begin{aligned}
\min_{{\bf x} \in {\bf \mathcal{X}}}~ & f({\bf x})\\
\textrm{s.t.}~ & x_i^{lb} \le x_i \le x_i^{ub}\\
\end{aligned}
\end{equation}

\noindent where, ${\bf x} \in \mathbb{R}^{n_x} \rightarrow \mathbb{R}$, and $x_i^{lb}$ and $x_i^{ub}$ are the lower and upper bounds of ${\bf x}$. As stated above, the GP serves as a surrogate function to represent the objective function $f({\bf x})$. The surrogate objective function can be defined by the mean value of the GP prediction given by $\hat{f}_{\mathcal{GP}}:=\mu_{\mathcal{GP}}$, or it can include an exploratory term that includes the GP variance to incentivise exploration,

\begin{equation}\label{eq:explGP}
    \hat{f}_{\mathcal{GP}}({\bf x}):=\mu_{\mathcal{GP}}({\bf x})-\alpha \sigma_{\mathcal{GP}}({\bf x},{\bf x}').
\end{equation}

\noindent Thus, the new function to be optimised is $\hat{f}({\bf x})_{\mathcal{GP}}$. An initial GP was built with the training data of inputs ${\bf x}$ and outputs ${\bf y}$. A non-linear optimisation approach is implemented to minimise $\hat{f}({\bf x})_{\mathcal{GP}}$ and obtain an estimation of the next point $x_\mathrm{new}$. The real function $f({\bf x})$ is evaluated at $x_\mathrm{new}$ and a new GP is built using the initial training set with updated values of $x_\mathrm{new}$ and $f(x_{ new})$. \cite{DelRio2019,DelRio2021} This process is done until a convergence criterion is achieved. The Bayesian optimisation with the GP algorithm is detailed below in Algorithm \ref{algorithm_1}.

\begin{algorithm}
\SetAlgoLined
\textbf{Initialisation:} Obtain $N$ initial observations ${\bf X}$ and their evaluations at objective function $f({\bf x})$. Build a GP with training set $\{{\bf X}, {\bf y}\}$ to produce $\hat{f}({\bf x})_{\mathcal{GP}}$. Set number of iterations $n_i$, set $i:=0$.\\
 \While{$i \le$ termination criteria:}{
  \begin{enumerate}
      \item Solve non-linear optimisation problem ${\bf x}_i^*:=\mathrm{argmin}_{{\bf x} \in \mathcal{X}} \hat{f}({\bf x})_{\mathcal{GP}}$
      \item Evaluate the objective function $f$ at point\\ ${\bf x}_i^*$ to obtain $y_i:=f({\bf x}_i^*) + \epsilon$
      \item Add the new values ${\bf x}_i^*$ and $y_i$ to the training\\ data of the GP
      \item Update the GP with the new available data\\ to model the surrogate function $\hat{f}({\bf x})_{\mathcal{GP}}$
      \item $i:= i + 1$
  \end{enumerate}
 }
 \caption{Bayesian optimisation with Gaussian Processes as surrogate function.}
 \label{algorithm_1}
\end{algorithm}

One of the main advantages of implementing this type of black-box optimisation method for producing electrode microstructures with user-specific properties is that GPs do not need a full understanding of the complex mechanisms that take place within the electrode microstructure. Compared to other surrogate models, such as artificial neural networks, GPs do not require large amounts of training data, and therefore, they are ideal for expensive black-box optimisation~\cite{Bradford2018,Rasmussen1997}. The coupling of a GP as a surrogate model of the trained generator, $G_{\theta^{(G)}}(\mathrm{\textbf{z}})$, with a Bayesian optimisation for the generation of an optimum electrode was implemented considering the generator latent space, ${\bf Z}$, as inputs, and the microstructural and transport properties of the electrode, ${\bf Y}$, as outputs. 

\section{Implementation in a real electrode microstructure}
\subsection{Closed-loop Generation-optimisation process} \label{sec:method_closed_loop}
We implemented a deep kernel Bayesian optimisation framework for the closed-loop generation-optimisation of microstructures with optimum user-specified properties. Our approach consists of three integrated steps:

\begin{enumerate}
\item Train a GAN to obtain the generator as a virtual representation of the microstructure parameterised by $\theta_\mathrm{GAN}$ and ${\bf z}$, as described in the authors' previous work~\cite{Gayon-Lombardo2020}.
\item Create a training set, $T=\{{\bf Z},{\bf Y}\}$, consisting of a latent space $({\bf Z})$ and its corresponding calculated microstructural and transport properties $({\bf Y})$ and use it to build a Gaussian Process. Here, the inputs were scaled to lie between $[0,1]$. Input scaling is a popular feature scaling procedure that has been shown to improve the prediction quality~\cite{Aksoy2001}.
\item Perform an iterative Bayesian optimisation process according to Algorithm~\ref{algorithm_1} to obtain the optimum latent space:\\
${\bf z}_i^*:=\mathrm{argmin}_{{\bf z} \in \mathcal{Z}} \hat{f}({\bf Z})_{\mathcal{GP}}$
\end{enumerate}

\subsubsection{GAN training} 
The GAN architecture used for training is defined in table~\ref{Table:Gan_architecture}, where both the discriminator and the generator are fully convolutional neural networks, as defined by~\cite{Radford2015}. The latent space of the Generator is given by a random normal distribution $z \sim \mathcal{N}(0,1)$, where $z \in \mathbb{R}^{n_z \times l \times l \times l}$ and $l = 4$, therefore the total size of the input vector ${\bf z}$ is 64. The discriminator is composed of five convolutional layers, each followed by a batch normalisation. The first four layers use a leaky rectified linear unit (LeakyReLU) activation function, and the last layer implements a sigmoid activation function. The generator is composed of five transposed convolutional layers \cite{Dumoulin2018}, which expand the spatial dimensions in each step. Each layer is followed by a batch normalisation, and all layers implement a ReLU activation function except for the last layer, which uses a Softmax function. The hyper-parameters that define the GAN architecture were chosen as detailed in~\cite{Gayon-Lombardo2020}.

\begin{table*}[ht]
\caption{Dimensionality of each layer in the GAN architecture for each porous material (layers, dimensions, optimiser, input image size, number of training samples)}
\begin{tabular}{c|c|c|c|c|c|c|c|c}
\toprule
\multirow{2}{*}{\textbf{Layer}} & \multirow{2}{*}{\textbf{Function}} & \textbf{Input} & \textbf{Output} & \textbf{Spatial} & \multirow{2}{*}{\textbf{Stride}} & \multirow{2}{*}{\textbf{Padding}} & \textbf{Batch} & \textbf{Activation} \\
 &  & \textbf{channels} & \textbf{channels} & \textbf{Kernel} & & & \textbf{normalisation} & \textbf{function} \\
\midrule
\multicolumn{9}{l}{\textit{Discriminator}} \\
\midrule
D1 & Conv3d & 3 & 16 & $4 \times 4 \times 4$ & 2 & 3 & Yes & LeakyReLU \\
D2 & Conv3d & 16 & 32 & $4 \times 4 \times 4$ & 2 & 2 & Yes & LeakyReLU \\
D3 & Conv3d & 32 & 64 & $4 \times 4 \times 4$ & 2 & 2 & Yes & LeakyReLU \\
D4 & Conv3d & 64 & 128 & $4 \times 4 \times 4$ & 2 & 2 & Yes & LeakyReLU \\
D5 & Conv3d & 128 & 1 & $6 \times 6 \times 6$ & 1 & 0 & No & Sigmoid \\ 
\midrule
\multicolumn{9}{l}{\textit{Generator}}\\
\midrule
G1 & ConvTransposed3d & 1 & 512 & $4 \times 4 \times 4$ & 2 & 2 & Yes & ReLU \\
G2 & ConvTransposed3d & 512 & 256 & $4 \times 4 \times 4$ & 2 & 2 & Yes & ReLU \\
G3 & ConvTransposed3d & 256 & 128 & $4 \times 4 \times 4$ & 2 & 2 & Yes & ReLU \\
G4 & ConvTransposed3d & 128 & 64 & $4 \times 4 \times 4$ & 2 & 2 & Yes & ReLU \\
G5 & ConvTransposed3d & 64 & 3 & $4 \times 4 \times 4$ & 2 & 3 & No & Softmax \\ 
\bottomrule
\end{tabular}
\label{Table:Gan_architecture}
\end{table*}

To overcome the instabilities that are commonly encountered during GAN training, a one-sided label smoothing stabilisation was implemented~\cite{Goodfellow2016}. This method reduces the label 1 corresponding to real images by a constant, $\varepsilon$, such that the new label has the value of $1$ – $\varepsilon$. For all cases, $\varepsilon$ has a value of 0.1. A ratio of network optimisation for the generator and discriminator was set to $2:1$, updating the generator twice while the discriminator once at each optimisation step. Stochastic gradient descent was implemented for learning using the ADAM optimiser~\cite{Kingma2017}. The momentum constants were $\beta_1\ =\ 0.5,\ \beta_2\ =\ 0.999$ and the learning rate was set to $2 \times 10^{-5}$. All simulations were performed on a GPU (Nvidia TITAN Xp), and the training process is limited to 72 epochs (\textit{ca.} 48 h).

\subsubsection{GP training} 
\begin{figure}[!htpb]
\centering
\includegraphics[width=0.5\columnwidth]{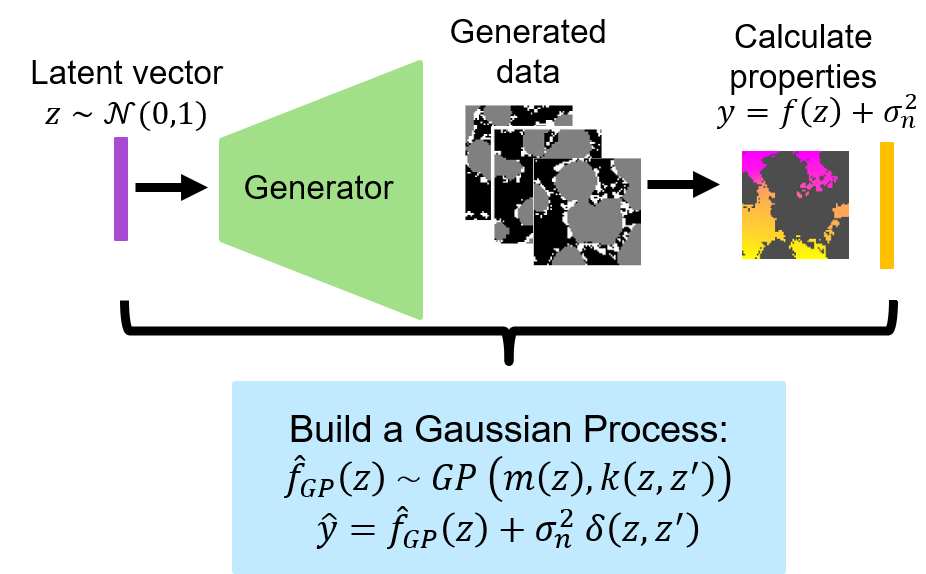}
\caption{Gaussian Process as surrogate model $\hat{f}({\bf x})_{\mathcal{GP}}$ to map the correlation between the latent space $\bf z$ and the estimated properties $\bf y$.}
\label{fig:GP}
\end{figure}

Figure \ref{fig:GP} summarises the process of creating a surrogate model $\hat{f}({\bf z})_{\mathcal{GP}}$ that can perform a mapping from the latent variables ${\bf z}$ as parameters of design, and the generated microstructure $G_\theta({\bf z})$ into the microstructural and transport properties ${\bf y} = f(G_\theta({\bf z}))$. 

An experiment design using Latin Hypercube Sampling (LHS) was performed, and a total of 50 64-dimensional vectors (\textit{\ie}, $1 \times 4 \times 4 \times 4$) were sampled in the $z$ space. Each of these vectors was used to generate a microstructure using the trained generator, and subsequently, its microstructural and transport properties were calculated using the open-source Matlab software TauFactor. The training set for the GP was formed by the 50 ${\bf z}$ vectors and their respective microstructural properties ${\bf y}$ as $T = \{{\bf Z},{\bf Y}\}$.

The GP was fully defined by the hyperparameters ${\bf \Theta} = [w_1, \dots, w_D, \sigma_f^2, \sigma_\epsilon^2]$. Since the hyperparameters were unknown \textit{a priori}, they were estimated by calculating the NLL of the joint PDF (Eq,~\ref{eq:NLL}). A multi-start search was performed by sampling five initial points determined by the LHS to avoid reaching a local minimum. A gradient-based optimisation was performed for each of these points using the bounded (Sequential Least Squares Programming) SLSQP algorithm~\cite{Kraft1985}, and the values with the minimum NLL were chosen as the fitted hyperparameters ${\bf \Theta^*}$.

\subsubsection{Bayesian optimisation} 
\begin{figure}[!htpb]
\centering
\includegraphics[width=0.5\columnwidth]{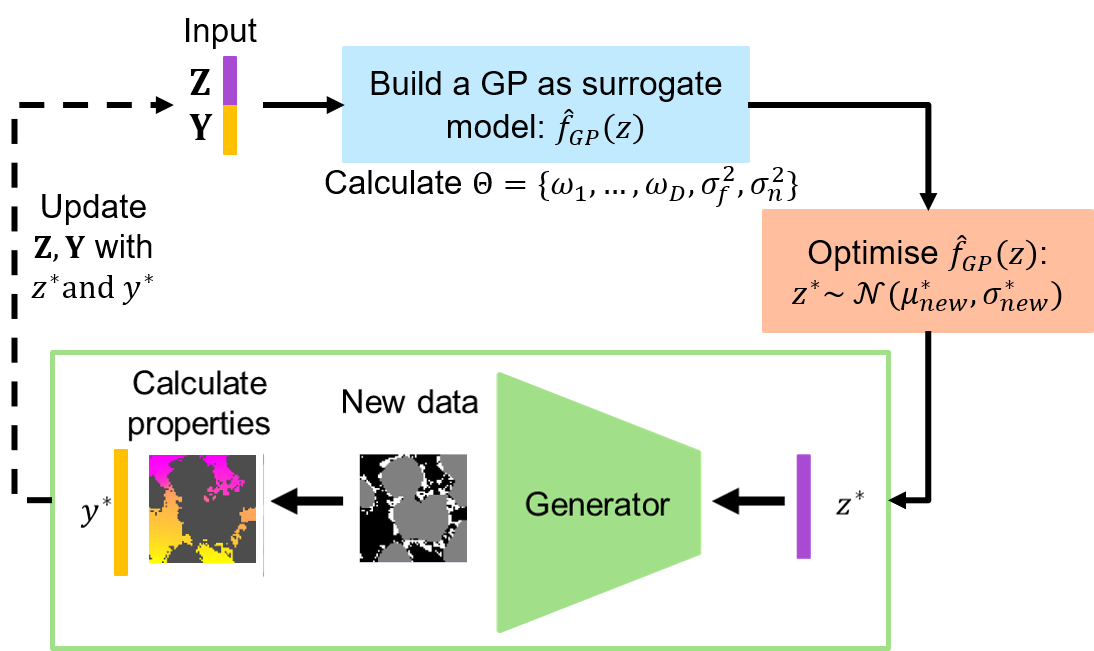}
\caption{Closed-loop generation-optimisation process. Bayesian optimisation algorithm to optimise the morphological and transport properties $\bf y$ of the generated microstructure as a function of the latent space $\bf z$ of the Generator.}
\label{fig:Bayesian}
\end{figure}

The integrated generation-optimisation closed-loop is shown in Figure~\ref{fig:Bayesian}.\\

\noindent The optimisation problem was defined by the objective function $f({\bf z})$, which represents the microstructural or transport property to be maximised as a function of the latent vector ${\bf z}$. Here, the analogous problem was implemented by the minimisation of the negative objective function, 

\begin{equation}
\begin{aligned}
{\bf z}^* =& \textrm{argmin}_{{\bf z} \in {\bf \mathcal{Z}}} -\hat{f}({\bf z}),\\
\textrm{s.t.}~& {\bf z}\in[-5,5]\\
\end{aligned}
\label{eq:optimisation}
\end{equation}

\noindent where, ${\bf z}$ was bounded between $[-5,5]$. The value of $f({\bf z})$ was estimated by inserting ${\bf z}$ into the trained generator, and calculating the microstructural or transport properties with a physics-based simulation using the open source software TauFactor~\citep{Cooper2016}. Since the microstructure-property problem is computationally expensive, an iterative sequential sampling was performed in which at each iteration a Gaussian Process is built to map $f(G({\bf z}))$. A multi-start gradient-based optimisation with the SLSQP algorithm was performed over the GP inference (Eq.~\ref{eq:mean_new}) in order to obtain the new value of ${\bf z}^*$ (Eq.~\ref{eq:optimisation}) \cite{Li2018, Bradford2018,Richardson2017,Kraft1985}. The surrogate objective function including an exploratory term was defined as stated in Eq. \ref{eq:explGP}. The parameter $\alpha$ was varied between 0 and 1.96, 

\begin{equation}
    \alpha = 1.96 \cdot \left(1.0- \frac{i}{i_\mathrm{tot}}\right),
    \label{eq:alpha}
\end{equation}

\noindent to guarantee a confidence interval of 95\% for the calculated values~\cite{Bradford2018}. Here, $i$ is the iteration number, and $i_\mathrm{tot}$ corresponds to the total number of iterations, which was set to 500 and used as the stopping criterion for each electrode property optimised. Once ${\bf z}^*$ was found at each iteration, ${\bf y}^*$ is calculated with a physics-based model, and the new values of $\{{\bf z},{\bf y}\}$ were added to the training set $T = \{{\bf Z},{\bf Y}\}$. The GP was then updated with the new values of ${\bf Z}$ and ${\bf Y}$ and the new hyperparameters ${\bf \Theta^*}$ were calculated.

\subsection{Case study: Optimisation of a 3D Li-ion cathode}

Lithium-ion batteries (LIBs) are one of the leading technologies for electrochemical energy storage,  particularly for electric vehicles, where they display large-scale adoption. As the demand for the electrification of transport continues to rise, the development of LIBs is expected to grow rapidly over the next decade. Still, there are a series of technological challenges related to electrodes' micros-scale morphology that need to be addressed to guarantee LIBs with high energy and power densities, long cycling life, and good reliability. While adopting new chemistries for LIB's active materials is one approach to improving performance, this study will instead focus on the optimisation of the microstructure of the porous electrodes. Electrodes within LIBs play an essential role as they constitute the main sites where the electrochemical reactions coupled with transport processes occur.
In order to optimise the microstructure of LIBs electrodes, the closed-loop generation-optimisation process described in section \ref{sec:method_closed_loop} was implemented. Specifically, this case study focused on a Li-ion cathode, which is composed of three phases: NMC particles, binder and pores.

\noindent The generator's training data are images of a lithium-ion battery cathode obtained from open-source nano-tomography data~\cite{Usseglio-Viretta2018}. These images had already been segmented to distinguish between three phases (grey, white and black):
\begin{enumerate}[label=(\alph*)]
    \item particles of a ceramic active material (nickel manganese cobalt oxide, NMC 532) \--- grey;
    \item a conductive organic binder (polymer with carbon black) \--- white; and
    \item pores \--- black.
\end{enumerate} 

\noindent Details of the sample preparation, imaging, reconstruction, and segmentation approaches were reported by~\cite{Usseglio-Viretta2018}. The cathode microstructural specifications are summarised in Table S1 in the supplementary information~\cite{Gayon-Lombardo2020}. A total of $N=13824$ overlapping sub-volumes were extracted from the original data set using a sampling function with a stride of eight voxels. The spatial dimension of these cropped volumes was selected based on the average size of the largest structural element (\ie, particle size) to guarantee that at least two of these elements could fit in each sub-volume~\cite{Blair1993}. The N extracted sub-volumes constituted the training set for a Generative Adversarial Network (GAN). The generator was previously trained as described in detail in the author's previous work  ~\cite{Gayon-Lombardo2020}.

\subsubsection{Optimisation cases}
An ideal Li-ion cathode would have a maximum relative diffusivity in the pore phase ($D_\textrm{rel,pore}$, dimensionless) with percolating paths that would enhance the liquid-state diffusion of Lithium ions. This relative diffusivity is a measure of the ease with which diffusive transport occurs with respect to the intrinsic diffusivity of the pores ($D_\mathrm{pore}$, \si{\m\squared\per\s}) and is correlated to the porosity ($\phi_\mathrm{pore}$, dimensionless) and tortuosity ($\tau_\mathrm{pore}$, dimensionless) properties of the cathode.  

\begin{equation}
    D_\textrm{rel,pore} = \frac{D_\textrm{eff,pore}}{D_\textrm{pore}} =  \frac{\phi_\mathrm{pore}}{\tau_\mathrm{pore}}
    \label{eq:Drel}
\end{equation}

\noindent The maximisation of $D_\textrm{rel,pore}$ along one of the three directions is of particular interest since the transport of Lithium ions is usually predominant along one direction (from the current collector to the membrane). Additionally, a maximum specific surface area of the NMC phase ($SSA_{\mathrm{NMC}}$, \si{\per\micro\m}) is desired to enhance the electrochemical conversion at the active sites and increase the utilisation of NMC particles. This volume SSA is defined as the ratio between the interfacial area of the NMC phase and pores ($A_\mathrm{NMC/pore}$, \si{\square\micro\m}) and the total volume of the cathode ($V$, \si{\cubic\micro\m}), 

\begin{equation}
    SSA_\mathrm{NMC} = \frac{A_\mathrm{NMC/pore}}{V}
    \label{eq:SSA}
\end{equation}

However, the microstructure and transport properties are correlated making the cathode optimisation a non-trivial problem that would require considering the interplay between these properties. For instance, an increase in $(D_\textrm{rel,pore})$ would directly lead to an increase in $\phi_\mathrm{pore}$, which will lead to a decrease in $(\phi_\mathrm{NMC})$, producing a lower energy density cathode. An “optimum microstructure” in terms of electrochemical performance would involve a trade-off between the available microstructural and transport properties. Nonetheless, a thorough analysis of microstructures with different properties, particularly, an incremental analysis varying one property while keeping the others constant, is of interest to understanding the effect of certain properties on the cathode performance. Here, we generate microstructures with a wide range of microstructural and transport properties to further produce databases of structures with specific desired properties. The following properties are maximised separately by defining a specific maximisation function for each case:
\begin{enumerate}[label=(\alph*)]
    \item Specific surface area of NMC phase $(SSA_\mathrm{NMC})$
	\item Relative diffusivity of the pore phase $(D_\textrm{rel,pore})$
	\item Relative diffusivity of the pore phase along the x direction $(D_\textrm{rel,pore,}x)$
	\item Relative diffusivity of the pore phase $(D_\textrm{rel,pore})$ and Specific surface area of NMC phase $(SSA_\mathrm{NMC})$ simultaneously.
	\item Specific surface area of the NMC phase $(SSA_\mathrm{NMC})$ constrained by a constant phase volume fraction of the NMC phase $\phi_\mathrm{NMC}$.
	\item Relative diffusivity of the pore phase $(D_\textrm{rel,pore})$ constrained by a constant porosity $\phi_\mathrm{pore}$.
\end{enumerate}

%%% RESULTS AND DISCUSSION %%%
\section{Results and Discussion}

As a first step, microstructural properties were maximised without constraints according to Eq.~\ref{eq:optimisation}. This was performed to prove the concept of a closed-loop generation-optimisation algorithm for various microstructural properties essential for electrode design (cases a, b and c). Subsequently, two microstructural properties were optimised (\ie $D_{\mathrm{rel,b}}$ and $SSA_\text{NMC}$) by constraining the secondary effect on the other properties (case e and f). Finally, a function is defined to generate microstructures with graded porosity along one direction, for microstructures of size $64^3$ and $128^3$ voxels. An analysis of the distribution of microstructural properties in the latent space of the generator is further performed to explore the existence of a correlation between these variables.

\subsection{Unconstrained optimisation} \label{sec:unconstrained}

\subsubsection{Specific Surface Area of the NMC phase} \label{sec:SSA}

The objective function to maximise the specific surface area of the NMC phase without constraints is defined as
\begin{equation}
    f({\bf z}) = SSA_\mathrm{NMC}.
\end{equation}

\noindent As previously stated, maximising the SSA of the active material, in this case, the NMC phase, is of interest to ensure a high utilisation of the active material and to enhance the electrochemical reaction.

\begin{figure*}[ht]
\centering
\includegraphics{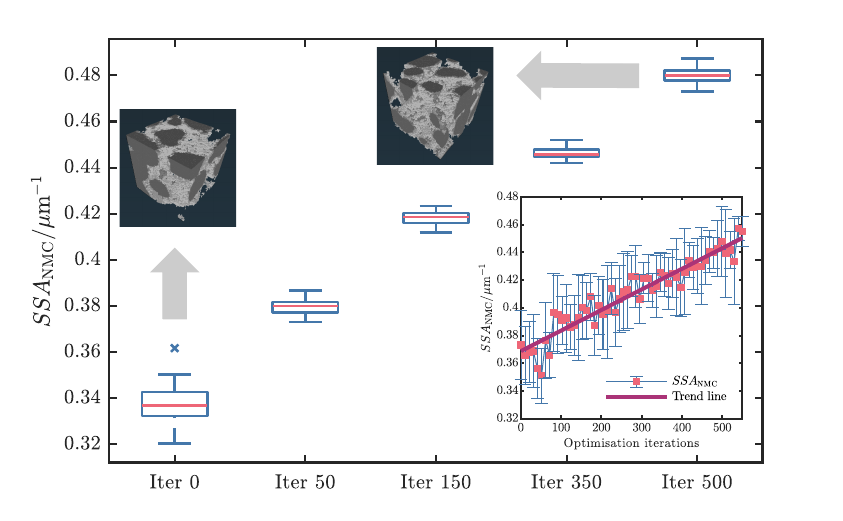}
\caption{Estimated SSA of 30 generated samples at points 0, 50, 150, 350 and 500 during the maximisation process. The inserted figure shows the complete unconstrained maximisation of the SSA of the NMC phase for 500 iterations.}
\label{fig:SSA}
\end{figure*}

The result of the maximisation process as a function of the number of iterations is shown in the inset of Figure~\ref{fig:SSA}. As the number of iterations increases, the $SSA_\mathrm{NMC}$ also increases without reaching a maximum since the optimisation is unconstrained. The iterative process is not smooth but rather oscillates significantly due to the existence of a search term when updating the Gaussian Process. This allows for the exploration of points within a trust region of 95\%. A set of microstructures was generated for each explored point throughout the optimisation process, as shown in Figure~\ref{fig:SSA}. In each case, 30 samples were generated from a latent vector ${\bf z}_i$, where $i$ corresponds to the iteration number. Each of these microstructures is unique, since ${\bf z}$ is different for each sample, but are visually similar given that their respective ${\bf z}$ is within the same region of the latent space, as will be shown in section~S4 in the supplementary information. Therefore, these microstructures also have similar properties. This indicates that a set of values in the latent space is correlated with the properties of the generated microstructure. When the samples of ${\bf z}$ are obtained from a normal PDF, the properties are in the same region as the GAN training set. However, microstructures with different properties are generated as the latent vector ${\bf z}$ is optimised and moves further from the normal distribution.

\begin{figure*}[!htpb]
\centering
\includegraphics[width=0.8\linewidth]{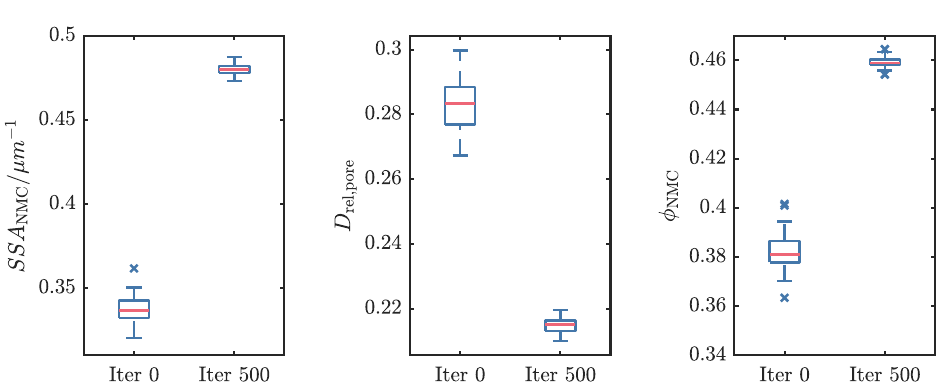}
\caption{Comparison between the estimated $SSA_\textrm{NMC}$, $D_\textrm{rel,pore}$ and $\phi_\textrm{NMC}$ at iteration 0 and 500 for 30 generated samples.}
\label{fig:SSA_Drel_VF}
\end{figure*}

Table~\ref{Table:Eqdiam_SSA} summarises the mean diameter and sphericity of particles within the microstructures at each iteration. The particle properties do not change when an unconstrained optimisation is performed, but the phase volume fraction of the NMC phase in the microstructures increases as shown in Figure~\ref{fig:SSA_Drel_VF}. As expected, since the NMC volume fraction increases, the pore volume fraction is reduced, leading to a reduction of the relative diffusivity in the pore phase, $D_{\text{rel,pore}}$. This change would result in a reduced volume of percolating paths, causing an increased transport resistance for Lithium ions through the liquid electrolyte. Therefore, an interplay between the effect of an enhanced $SSA_\mathrm{NMC}$ and a reduced $D_{\text{rel,pore}}$ must be considered in the electrode design and will be discussed in section~\ref{sec:Drel_SSA}. Mathematical correlations between the $SSA_\textrm{NMC}$ and $\phi_\textrm{pore}$ based on idealised particle systems with a homogeneous particle diameter have been previously proposed~\citep{Rabbani2014, Suthar2015}, however, their applicability depends on the type of microstructure. Figure S1 in the supplementary information compares the SSA as a function of the porosity for different correlations to the SSA calculated over the generated microstructures. Even though all SSAs display a negative slope with respect to porosity, the SSA is typically underestimated by these correlations, with the ideal packed particles correlation giving the closest results to the generated data. Nonetheless, the variability of the calculated SSA at a micro-scale level due to other microstructural factors apart from the porosity is not explained. Correlations' validity can be assessed using large amounts of non-idealised synthetic microstructures displaying different properties.

\begin{table}[ht]
\caption{Particle properties of microstructures samples during $SSA_\textrm{NMC}$ unconstrained maximisation.}
\begin{tabular}{c| c| c } 
\toprule
\textbf{Iteration} & \textbf{Equivalent diameter $/\mu\textrm{m}$} & \textbf{Sphericity} \\
\midrule
0	& 6.45	& 0.85 \\
100	& 5.51	& 0.87 \\
250	& 5.15	& 0.87 \\
350	& 5.57	& 0.88 \\
500	&6.25	& 0.87 \\
\bottomrule
\end{tabular}
\label{Table:Eqdiam_SSA}
\end{table}

\subsubsection{Relative diffusivity of the pore phase} \label{sec:Drel}

As described previously, high-power Li-ion electrodes would enable the fast transport of Lithium ions through the liquid electrolyte in the porous phase. Thus, a maximum relative diffusivity in the porous phase is desired to enhance ionic transport. To study this an unconstrained maximisation problem was defined by the objective function:

\begin{equation}
    f({\bf z}) = D_\text{rel,pore} = \frac{\phi_\text{pore}}{\tau_\text{pore}}.
    \label{eq:Drel_b}
\end{equation}

\begin{figure*}[ht!]
\centering
\includegraphics[trim= 0cm 0.5cm 0cm 0cm, clip]{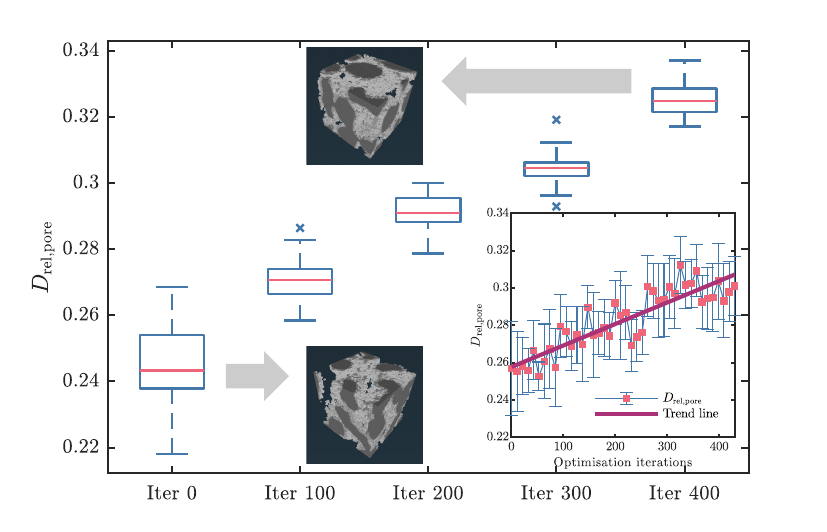}
\caption{Estimated $D_\textrm{rel,pore}$  of 30 generated samples at each 100 points during the maximisation process. The inserted figure shows the complete unconstrained maximisation of $D_\textrm{rel,pore}$ for 400 iterations.}
\label{fig:Drel}
\end{figure*}

\noindent The result of the unconstrained maximisation of $D_\textrm{rel,pore}$ of the pore phase is shown in the inset of Figure~\ref{fig:Drel}.

\begin{figure*}[ht!]
\centering
\includegraphics[width=1\linewidth]{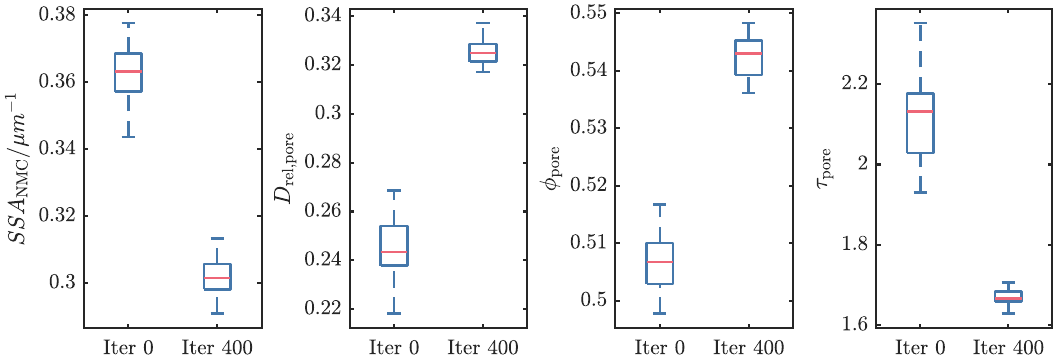}
\caption{Comparison of the estimated $SSA_\textrm{NMC}$, $D_\textrm{rel,pore}$, $\phi_\text{pore}$, and $\tau_\textrm{pore}$ for iteration 0 and 400 of the unconstrained maximisation process of $D_\textrm{rel,pore}$.}
\label{fig:Drel_compare}
\end{figure*}

The maximisation process changes the transport properties, however, the resulting microstructures are visually indistinguishable from the training data as shown in the inserted 3D reconstructions at iterations 0 and 400. This is a property attributed to the generator since it is trained to recreate synthetic microstructures with the same probability distribution function as the real tomographic data. Figure~\ref{fig:Drel_compare} compares the $SSA_\textrm{NMC}$ of the initial and optimised microstructures, which decreases as $D_\textrm{rel,pore}$ increased. This inverse correlation is expected as explained in the previous section. Additionally, a maximisation of $D_\textrm{rel,pore}$ could be purely attributed to an increase in the pore volume fraction, however, as it can be seen in Figure \ref{fig:Drel_compare}, not only is $\phi_\text{pore}$ increased but also the tortuosity factor $\tau$ is reduced. This indicates that the configuration of the microstructure is manipulated through the optimisation of the latent space to enable the existence of less tortuous paths that reduce the material resistance towards an incoming Li diffusive flow.

These results show the existence of a strong correlation between the microstructural and transport properties and therefore highlight the need to implement constraints during the optimisation process. By considering the trade-off between the $SSA$ of the NMC phase and the $D_\textrm{rel}$ of the pore phase, a question arises, whether an equilibrium exists for maximising these two properties simultaneously, thus enhancing both the electrochemical reaction and the fast transport of Li ions.

This optimisation case was implemented to generate cathode volumes of $64^3$ voxels. However, volumes large enough are needed in order to simulate representative cathodes. Based on the fully convolutional architecture of the generator, microstructures of any size can be generated by increasing the size of the latent space \cite{Gayon-Lombardo2020}. To demonstrate the implementation of this method for larger microstructures, a maximisation of the effective diffusivity in the pore phase is implemented for a volume of $128^3$ voxels. The objective function is defined as Eq. \ref{eq:Drel_b}. The results of the maximisation process for a maximum of 800 iterations are shown in Figure S2 in the supplementary information. These results prove the effectiveness of implementing the proposed closed-loop optimisation process for a large microstructure achieving a visually realistic microstructure with increased relative diffusivity in the pore phase. Moreover, a larger set of microstructures of the same size with different tailored properties can be generated based on the optimisation process of the latent space. Thus, knowing the values of the vectors $\bf z$ which are correlated to their respective values of $D_\textrm{rel,pore}$ is equivalent to encoding the large microstructures into a 64-digit "code" where the $128^3$ microstructure can be rapidly generated ($\sim$\SI{3}{\s}) with the generator. This allows a large amount of three-dimensional data to be saved in a computationally inexpensive manner.

\subsubsection{Relative diffusivity of the pore phase and Specific Surface Area of the NMC phase} \label{sec:Drel_SSA}

As shown in sections \ref{sec:SSA} and \ref{sec:Drel}, the microstructure can be manipulated to maximise the microstructural and transport properties that are known to increase cell performance. However, these properties are not independent and are strongly correlated. Mathematical functions to correlate the effect of tortuosity factor with porosity and $SSA$ with porosity have been previously proposed, however these correlations present deviations for different microstructures (as shown in section \ref{sec:SSA})~\cite{Cooper2017}. Therefore, in order to understand the effect of maximising mutually correlated properties, an objective function must be defined to optimise the desired microstructural and transport properties while constraining the values of the correlated properties. For the purpose of this work, to obtain a microstructure with maximum values of $SSA_\textrm{NMC}$ and $D_\textrm{rel,pore}$, both properties must be optimised simultaneously in the definition of $f(z)$. The objective function was defined as:
\begin{equation}
    f({\bf z}) = \beta \cdot D_\text{rel,pore,norm} + \gamma \cdot SSA_\text{NMC,norm},
\end{equation}

\noindent where the normalised properties were defined by considering a property range,
\begin{equation}
    D_\text{rel,pore,norm} = \frac{D_\text{rel,pore}}{D_\text{rel,pore,range}},
\end{equation}

\begin{equation}
    D_\text{rel,pore,range} = D_\text{rel,pore,max} - D_\text{rel,pore,min},
\end{equation}

\begin{equation}
    SSA_\text{NMC,norm} = \frac{SSA_\text{NMC}}{SSA_\text{NMC,range}},
\end{equation}

\begin{equation}
    SSA_\text{NMC,range} = SSA_\text{NMC,max} - SSA_\text{NMC,min}.
\end{equation}

\noindent By definition, $\beta, \gamma \in [0, \dots, 1]$ are coefficients that determine the weight of each property in the objective function. These two coefficients are related to each other by,

\begin{equation}
    \beta = 1 - \gamma.
    \label{eq:beta}
\end{equation}

\begin{figure*}[ht!]
\centering
\includegraphics[width=0.8\linewidth]{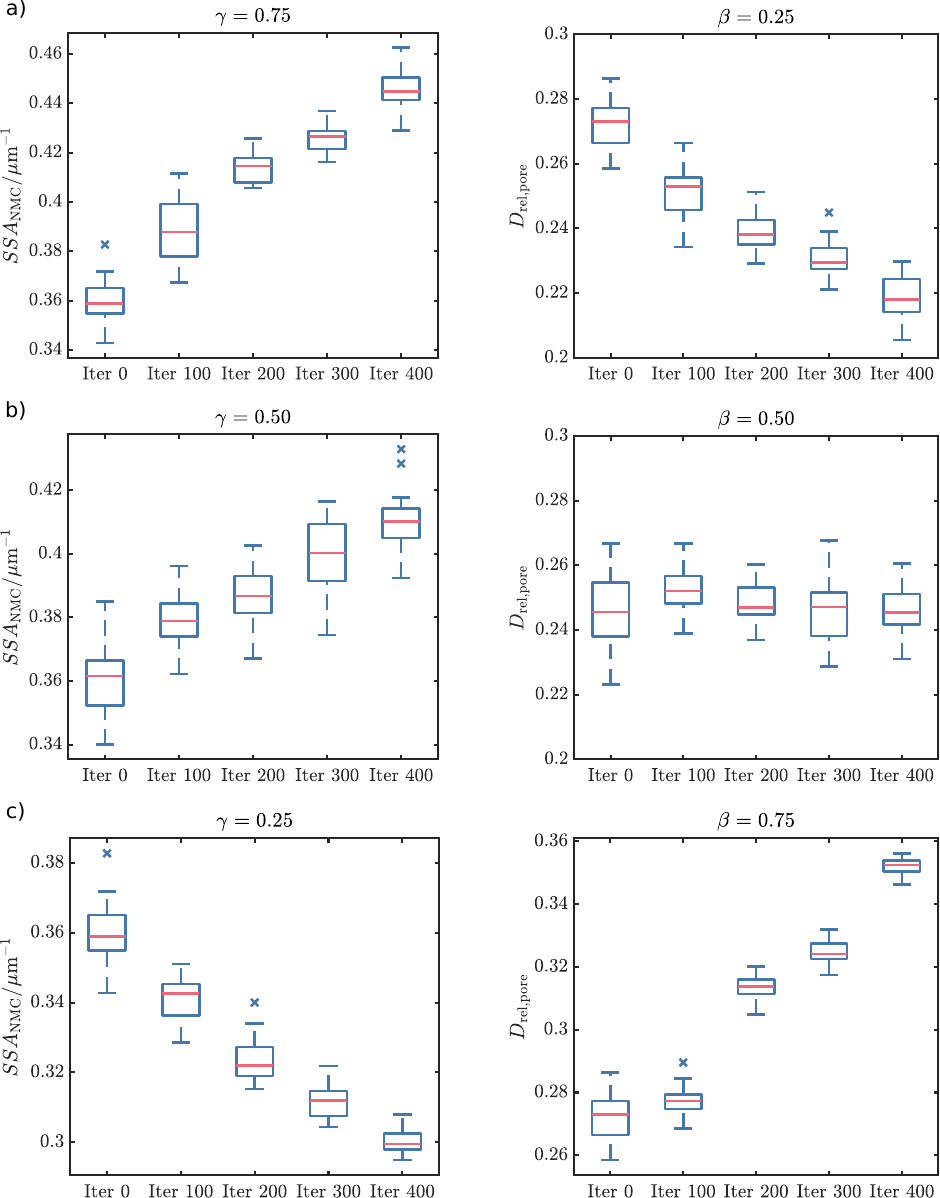}
\caption{Results of estimated $SSA_\textrm{NMC}$ and $D_\textrm{rel,pore}$ as a function of the iterations number for different values of $\beta$: a) $\beta = 0.25$, b) $\beta = 0.5$, c) $\beta = 0.75$}
\label{fig:SSA_Drel}
\end{figure*}

A sensitivity analysis was performed to evaluate the impact of each property in the objective function by varying the values of $\beta$ and $\gamma$ between 0.25, 0.5, and 0.75. Figure \ref{fig:SSA_Drel} presents the optimisation result for each value of $\gamma$ after $500$ iterations. 

Based on the assumption that both properties contribute equally to the maximisation of the objective function, a microstructure with maximum $SSA_\textrm{NMC}$ and $D_\text{rel,pore}$ simultaneously would be obtained by assigning the value of $\gamma = 0.5$. However, in Figure \ref{fig:SSA_Drel} it is seen that a value of $\gamma = 0.5$ leads to an increase in $SSA_\textrm{NMC}$ (positive slope) while keeping the values of $D_\textrm{rel,pore}$ almost constant (slope approx. 0). By reducing the value of $\gamma$ to $0.25$, the slope of $D_\textrm{rel,pore}$ becomes positive and the slope of $SSA_\textrm{NMC}$ becomes negative. The fact that the slope of $SSA_\textrm{NMC}$ is inverted indicates that a maximum in $f({\bf z})$ where both properties present a positive slope would be found at a value of $\gamma$ between $0.5$ and $0.25$. This analysis shows that the objective function is more sensitive towards a variation in the $SSA_\textrm{NMC}$ than a variation in $D_\textrm{rel,pore}$, and therefore, a small increase in the coefficient of $SSA_\textrm{NMC}$ leads to a significant increase in the objective function. This also indicates that there is a more direct correlation between the latent space of the generator and the $SSA_\textrm{NMC}$ than with other properties. This is important to consider for future work where the latent space could be implemented as parameters of the design of optimum microstructure.

When analysing the other two cases, where $\gamma = 0.75$ and $0.25$, it is clear that even though the correlation between the two properties is inverse, by manipulating the coefficient of each property in the objective function (\ie $\beta$ and $\gamma$) it is possible to obtain microstructures with a large improvement in a specific property (maximisation with positive slope) while constraining the decrease in the values of the correlated property. These results highlight the importance of considering the trade-off between property values when optimising and designing electrode microstructures.

\subsection{Constrained optimisation}\label{sec:constrained}

\subsubsection{Specific Surface Area of the NMC phase constrained by the NMC volume fraction}

One pathway to produce a cathode with maximum $SSA_\textrm{NMC}$ is to increase its NMC volume fraction ($\phi_\mathrm{NMC}$). Although this is theoretically achievable, as previously shown, an increase in the NMC loading leads to a decrease in the porosity, and therefore a decrease in the $D_\textrm{rel,pore}$. For this reason, it is desired to optimise the accessible capacity for a fixed total NMC loading (\ie, fixed volume fraction of NMC material). Based on this, a maximisation of $SSA_\textrm{NMC}$ must be constrained to maintaining the volume fraction of the NMC material constant. Thus, the objective function to optimise was defined by 

\begin{equation}
    f(z) = \frac{SSA_\text{NMC}}{SSA_\text{NMC,range}} - \frac{\textrm{RMSE}(\phi_\textrm{NMC} - \phi_\textrm{NMC,mean})}{\phi_\textrm{NMC,range}}.
    \label{eq:SSA_const}
\end{equation}

\noindent Where the second term on the right-hand side of the equation corresponds to a penalisation term to keep the volume fraction constant, defined as

\begin{equation}
    \textrm{RMSE} = \sqrt{\frac{\sum_{i=1}^N \left(\phi_{\textrm{NMC}_i} - \phi_{\textrm{NMC}_{mean}} \right)^2}{N} }.
    \label{eq:RMSE_vf}
\end{equation}

\noindent Equation \ref{eq:RMSE_vf} corresponds to the Root-Mean-Squared Error (RMSE) of $\phi_{\textrm{NMC}_i}$ for $i = \{1, \dots, N\}$ where $N$ is the total number of samples, and $\phi_{\textrm{NMC}_{mean}}$ corresponds to the mean value of $\phi_\textrm{NMC}$ in the training set.

Previous authors have proposed an optimisation of the microstructure by decreasing the particle size to enable a higher specific surface area. These works however are based on idealised representations of the microstructure constituted by spherical particles whose diameter can be decreased according to an objective function. In this work the particle size is not a target of the objective function, but rather the resulting SSA. The optimisation is performed directly over the architecture of the microstructure defined by the latent space, without targeting a particular property.

\begin{figure*}[ht!]
\centering
\includegraphics{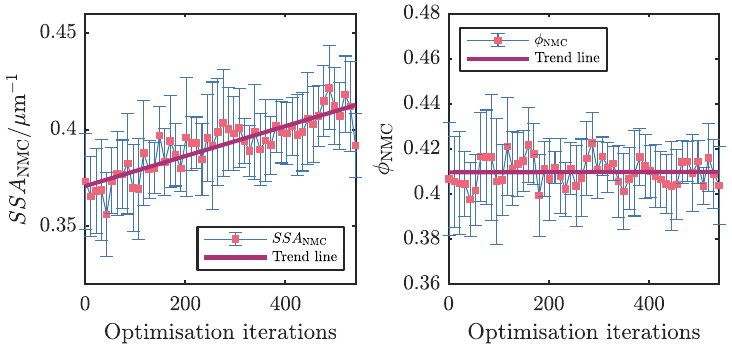}
\caption{Results of estimated $SSA_\textrm{NMC}$ and $\phi_\textrm{NMC}$ as a function of the iterations number for the $SSA_\textrm{NMC}$ maximisation process constrained by a constant $\phi_\textrm{NMC}$. The results show a confidence interval of 95\%.}
\label{fig:SSA_VF}
\end{figure*}

The results of the maximisation of the $SSA_\textrm{NMC}$ constraining the NMC volume fraction are shown in Figure \ref{fig:SSA_VF}. These results show an increase in the $SSA_\textrm{NMC}$ while the values of the $\phi_\textrm{NMC}$ remain constant. By comparing the estimated average particle size or equivalent diameter and the sphericity of the initial microstructure with the optimised microstructure reported in Table \ref{Table:Eqdiam_SSA_vfix}, it is shown that the mean particle size does not vary significantly. This implies that the maximisation of the $SSA_\textrm{NMC}$ is not always related to the particle size or sphericity, but can be a result of a redistribution of the active particles in the given space, or a change in the roughness of the outer NMC surface that can lead to an increase in specific surface area.

\begin{table}[h]
\caption{Equivalent diameter and sphericity of Li-ion cathode microstructures samples during $SSA_\textrm{NMC}$ maximisation constraining $\phi_\mathrm{NMC}$.}
\begin{tabular}{c| c| c} 
\toprule
\textbf{Iteration} & \textbf{Equivalent diameter $/\mu\textrm{m}$} &  \textbf{Sphericity} \\
\midrule
0 & 5.31 & 0.88  \\ 
100 & 6.13	& 0.85 \\
200 & 7.03	& 0.81 \\
300 & 5.73	& 0.83 \\
500 & 6.43	& 0.81 \\
\bottomrule
\end{tabular}
\label{Table:Eqdiam_SSA_vfix}
\end{table}

\subsubsection{Relative diffusivity of the pore phase constrained by the pore phase volume fraction}

As previously stated, an unconstrained maximisation of the relative diffusivity in the pore phase ($D_\textrm{rel,pore}$) leads to an increase in the cathode porosity, which subsequently decreases the loading of active material. In this respect, a maximisation of the $D_\textrm{rel,pore}$ needs to be constrained by keeping the porosity constant. This objective function is defined as

\begin{equation}
    f(z) = \frac{D_\text{rel,pore}}{D_\text{rel,pore,range}} - \frac{\textrm{RMSE}(\phi_\text{pore} - \phi_\text{pore,mean})}{\phi_\text{pore,range}}.
    \label{eq:Drel_const}
\end{equation}

\noindent Where the second term on the right-hand side of the equation corresponds to the penalisation term involving the RMSE of the porosity, $\phi_\text{pore}$, calculated over the $N$ generated samples.

\begin{figure*}[ht!]
\centering
\includegraphics[width=\textwidth]{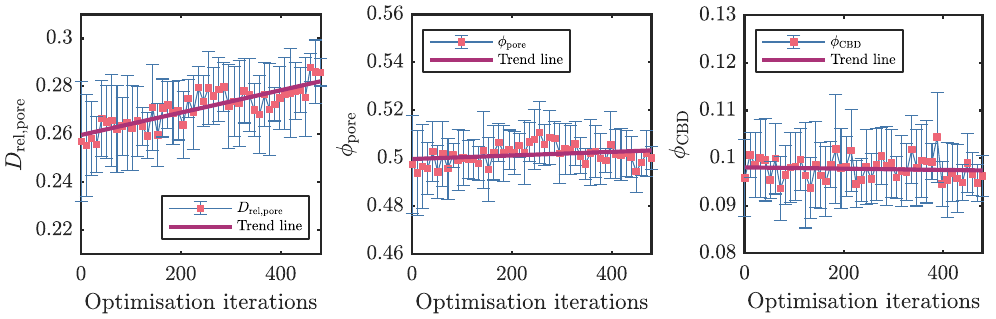}
\caption{Results of estimated $D_\textrm{rel,pore}$, $\phi_\text{pore}$ and $\phi_\text{CBD}$ as a function of the iterations number for the $D_\textrm{rel,pore}$ maximisation process constrained by a constant $\phi_\text{pore}$. The results show a confidence interval of 95\%.}
\label{fig:Drel_VF}
\end{figure*}

The results of the increase in $D_\textrm{rel,pore}$ as a function of the number of iterations are shown in Figure \ref{fig:Drel_VF}. This shows that an increase in the $D_\textrm{rel,pore}$ is achieved after 500 iterations, while the $\phi_\text{pore}$ is kept constant. In comparing the constrained and unconstrained maximisation, the positive slope of the constrained iterative process is not as steep as the one for the unconstrained maximisation of $D_\textrm{rel,pore}$, shown in Figure \ref{fig:Drel}. This proves that exist a tight correlation between $D_\textrm{rel,pore}$ and $\phi_\text{pore}$. Given that the $\phi_\text{pore}$ is kept constant for the constraint optimisation, it is rationalised that the maximisation of the relative diffusivity is achieved by a decrease in the electrode tortuosity, $\tau_\mathrm{pore}$. This means that the latent space optimisation allows a restructuring of the microstructure which results in the opening of flow paths that enable the transport of lithium ions. Additionally, the effect of the carbon-binder domain (CBD) during the constrained optimisation was analysed. The purpose of the CBD is to provide mechanical integrity to the electrode and conduct electrons. Thus, a change in the CBD load would directly impact the electronic conductivity of the cathode. From Figure \ref{fig:Drel_VF} it is seen that a constraint in the porosity also leads to a fixed volume fraction of the CBD, and by definition of volume fraction, $\phi_\mathrm{pore} + \phi_\mathrm{NMC} + \phi_\mathrm{CBD} = 1$, the loading of active material, $\phi_\mathrm{NMC}$, is also kept constant.

\subsubsection{Relative diffusivity of the pore phase along the x-direction}

A characterisation of the original microstructure shown elsewhere \cite{Gayon-Lombardo2020} indicated that the microstructure is isotropic, and therefore, all its properties are statistically the same along the three directions (through-plane and in-plane). Nonetheless, it is known that the transport of lithium ions during the battery cycling is predominant along the through-plane direction ($x$-direction), which consists of the transport of lithium ions from the membrane to the current collector. Based on this, an improved relative diffusivity in the pore phase is desired along this direction of transport, and therefore, a maximisation of the $D_{\text{rel,pore},x}$ is considered as the objective function, defined as

\begin{equation}
    f({\bf z}) = D_{\text{rel,pore},x} = \frac{\phi_{\text{pore},x}}{\tau_x}.
    \label{eq:Drelb}
\end{equation}

\noindent This equation does not constraint the increase of the $D_\textrm{rel,pore}$ along the other two directions, nonetheless it favours an increase of this property in the direction to be maximised (\ie, the $x$-direction).

\begin{figure*}[ht!]
\centering
\includegraphics[width=\textwidth]{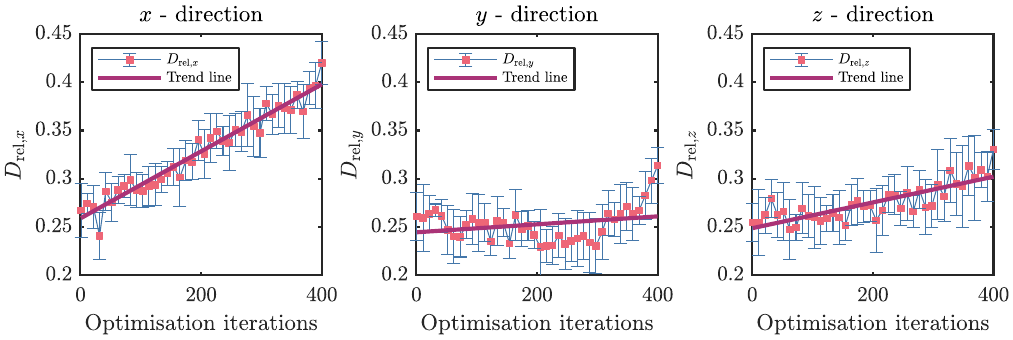}
\caption{Results of the estimated $D_\textrm{rel,pore}$ as a function of the iterations number for the three directions $x$, $y$ and $z$ for the unconstrained maximisation of  $D_{\text{rel,pore},x}$. The results show a confidence interval of 95\%.}
\label{fig:Drel_dir_unc}
\end{figure*}

\begin{figure*}[ht!]
\centering
\includegraphics[width=1\linewidth]{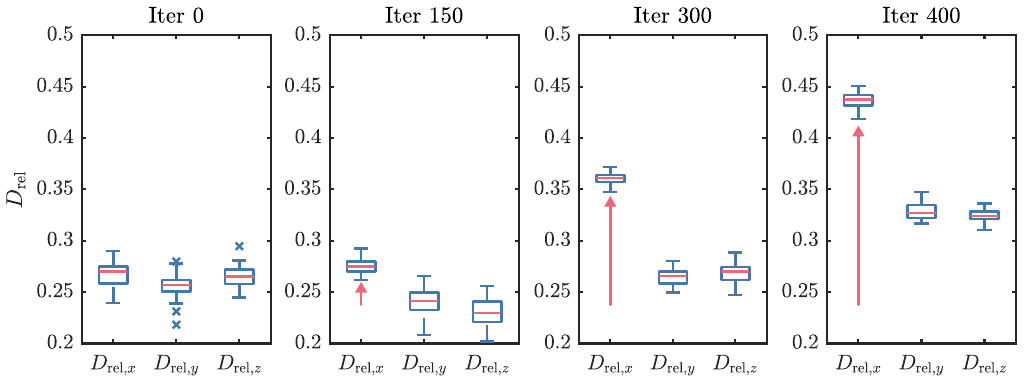}
\caption{Results of estimated $D_\textrm{rel,pore}$ of 30 microstructure samples generated at four points during the unconstrained maximisation process of  $D_{\text{rel,pore},x}$.}
\label{fig:Drel_dir_unc_micr}
\end{figure*}

Figures \ref{fig:Drel_dir_unc} and \ref{fig:Drel_dir_unc_micr} summarise the results of maximising Eq.~\ref{eq:Drelb}. The $D_\textrm{rel,pore}$ along direction $y$ and $z$ remains virtually constant throughout all iterations, until iteration 200, where a further increase in the $D_{\text{rel,pore},x}$ imposes an increase in $D_{\text{rel,pore},y}$ and $D_{\text{rel,pore},z}$. In order to constrain the $D_\textrm{rel,pore}$ to be constant along $y$ and $z$ directions, a penalisation term for each direction is added to the objective function, as given by

\begin{equation}
\begin{aligned}
    f({\bf z}) =& D_{\text{rel,pore},x} - \textrm{RMSE}\left(D_{\text{rel,pore},y},D_{\text{rel,pore},y_{mean}} \right) \\
    &- \textrm{RMSE}\left(D_{\text{rel,pore},z},D_{\text{rel,pore},z_{mean}} \right).
    \label{eq:Drel_dir_const}
\end{aligned}
\end{equation}

\noindent Where the RMSE of $D_{\text{rel,pore},y}$ (second term on the right hand side of Eq. \ref{eq:Drel_dir_const}) is given by
\begin{equation}
    \textrm{RMSE} = \sqrt{\frac{\sum_{i=1}^N \left(D_{\text{rel,pore},y_i} - D_{\text{rel,pore},y_{mean}} \right)^2}{N} },
    \label{eq:RMSE}
\end{equation}

\noindent for $i = \{1, \dots, N\}$ where $N$ constitutes the total number of samples, and $D_{\text{rel,pore},y_{mean}}$ corresponds to the mean value of $D_{\text{rel,pore},y}$ in the training set. An analogous equation to Eq. \ref{eq:RMSE} was also applicable to $D_{\text{rel,pore},z}$ (third term on the right hand side of Eq. \ref{eq:Drel_dir_const}).

\begin{figure*}[ht!]
\centering
\includegraphics[width=\textwidth]{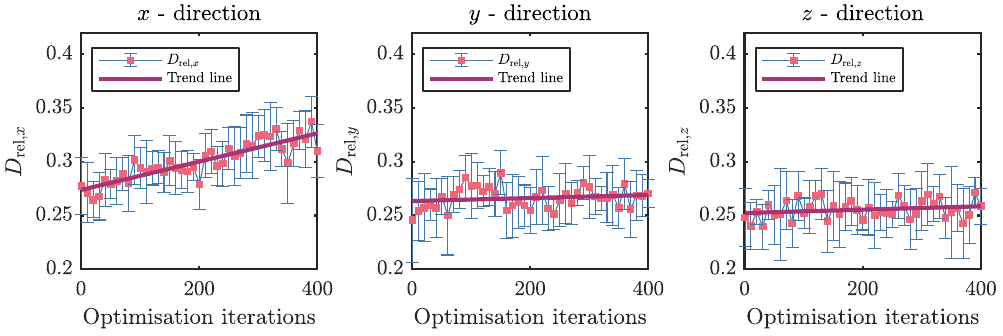}
\caption{Results of the estimated $D_\textrm{rel,pore}$ as a function of the iterations number for the three directions $x$, $y$ and $z$ for the maximisation of  $D_{\text{rel,pore},x}$ constrained by a constant value of $D_{\text{rel,pore},y}$ and $D_{\text{rel,pore},z}$. The results show a confidence interval of 95\%}
\label{fig:Drel_dir}
\end{figure*}

\begin{figure*}[ht!]
\centering
\includegraphics[width=1\linewidth]{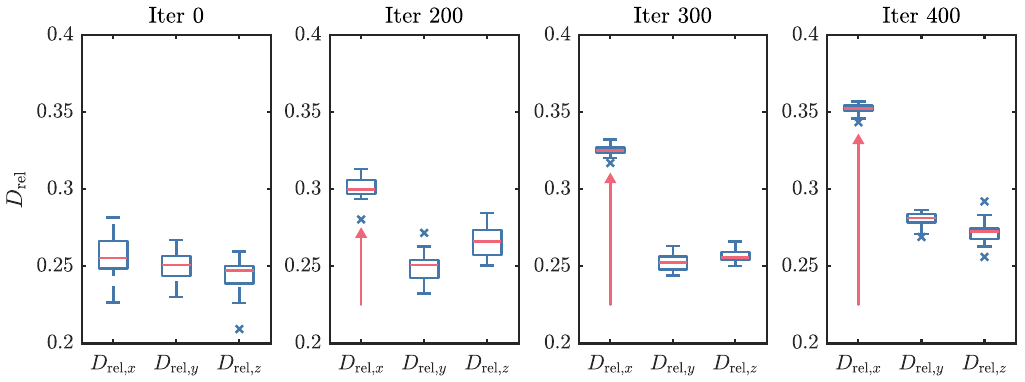}
\caption{Results of estimated $D_\textrm{rel,pore}$ of 30 microstructure samples generated at four points during the maximisation process of $D_{\text{rel,pore},x}$ constrained by a constant value of $D_{\text{rel,pore},y}$ and $D_{\text{rel,pore},z}$.}
\label{fig:Drel_dir_micr}
\end{figure*}

The results in Figure \ref{fig:Drel_dir} and \ref{fig:Drel_dir_micr} show that by imposing two penalisation terms in the objective function corresponding, the maximisation of $D_{\text{rel,pore},x}$ is achieved while $D_{\text{rel,pore},y}$ and $D_{\text{rel,pore},z}$ are kept constant. In comparing Figure \ref{fig:Drel_dir_unc} and Figure \ref{fig:Drel_dir}, it is clear that the increase in $D_{\text{rel,pore},x}$ is not as significant when the penalisation terms were added in the other two directions. This proves the tight correlation between the properties along each direction; however, it also shows that the directionality of each property can be optimised independently and does not necessarily change the properties along the other directions.

\vspace{3\baselineskip}

Up until now, this work has proved that an optimisation of the latent vector $({\bf z})$ of the generator as parameters of design can lead to the generation of microstructures $(G({\bf z}))$ with customised properties. In order to explore if a correlation exists between the latent space and the various microstructural properties, the 64-dimensional latent space was reduced into two principal components and visualised against the microstructural properties.

Figure S3 in the supplementary information summarises the results of squeezing the 64-dimensional latent space into two principal components. Based on this test, it is clear that the two principal components of the latent space are related to the electrode's microstructural properties. Moreover, this results in the creation of a space where the neighbouring principal components contain similar microstructural properties.

\subsection{Graded porosity optimisation} \label{sec:graded}

Improved battery performance has been previously achieved through the design of electrodes with graded porosity and graded particle size distribution \cite{Lu2020}. These graded properties have been achieved through a direct manipulation of the tomographic data, this is by changing a specific set of voxels to increase or decrease the phase volume fraction. As an alternative method, this work proposes the implementation of the closed-loop generation optimisation approach for the generation of new microstructure with graded porosity or particle distribution. By implementing a target volume fraction at the inlet and outlet walls and defining a linear space between these points of the same size as the electrode length, an objective function can be defined to fit the volume fraction of each image to the target volume fraction. This function is defined as

\begin{equation}
    f({\bf z}) = \textrm{RSME}(\phi_{p,j},\phi_{{\textrm{linspace}},j}),
    \label{eq:graded}
\end{equation}

\noindent where the RMSE is given by

\begin{equation}
    \textrm{RMSE} = \sqrt{\frac{\sum_{j=1}^{m=64} \left(\phi_{p,j} - \phi_{\textrm{linspace},j} \right)^2}{m} },
    \label{eq:RMSE_VF}
\end{equation}

\noindent and $p$ refers to the phase (\ie, pore, NMC or CBD), and $j$ corresponds to the size of the electrode length. The total length in voxels was 64, which corresponds to the number of stacked 2D images obtained from the tomographic data.

\begin{figure*}[ht!]
\centering
\includegraphics[width=0.75\linewidth]{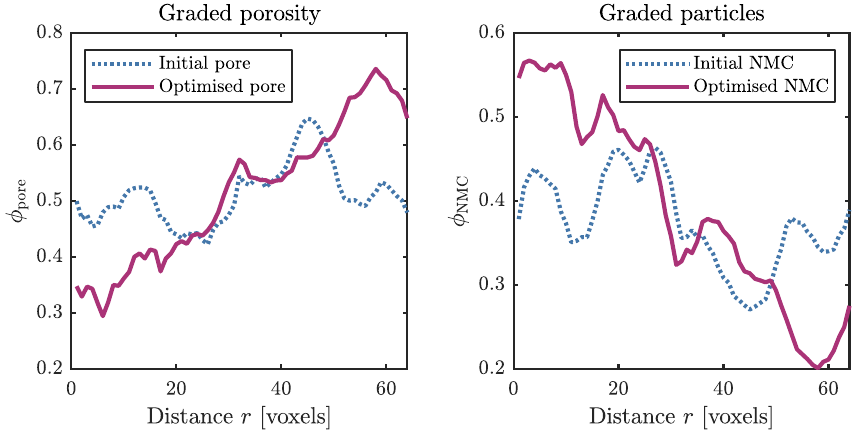}
\caption{Unconstrained graded optimisation of the volume fraction of pore (porosity) and NMC as a function of the electrode length along the direction to be optimised after 30 iterations.}
\label{fig:Graded}
\end{figure*}

Figure \ref{fig:Graded} shows the initial and optimised volume fraction of the pore and NMC phases as a function of the electrode length. It can be seen that the initial image contained a random distribution of the phases' volume fraction. After 30 iterations, the resulting volume fraction is graded along a positive or negative slope in the direction of the corresponding values of the linear space. These results show that an optimisation of the $\bf z$ space not only leads to a change in the microstructural properties as homogenised values but can also provide directionality to the properties. This can be implemented for any property by defining an objective function similar to Eq. \ref{eq:graded}.

%%% CONCLUSION %%%
\section{Conclusions}

In the field of electrode design and optimisation for energy storage applications, this work presents an innovative methodology that leverages generative models and Bayesian optimisation to design electrode microstructures with specific, user-defined properties. This work introduces a closed-loop generation-optimisation process to enable the generation of synthetic microstructure with optimum properties and implements this method as a case study for the design of improved cathode microstructures of Lithium-ion batteries.

We demonstrated the implementation of a deep kernel learning model, where a GP regression serves as a surrogate function that maps the latent space of a trained Generative Adversarial Network (GAN) generator to the microstructural and transport properties of the generated microstructure. The use of a GAN's trained generator reduces information loss and preserves microstructural characteristics throughout the closed-loop optimisation process. Moreover, implementing a deep kernel Bayesian optimisation allows for the reduction of the number of design evaluations, thus decreasing computational cost. This efficiency arises from the Bayesian optimisation's non-gradient nature, enabling a more flexible, gradient-free search for an optimal latent space to produce microstructures with targeted properties.

This work also demonstrated that the closed-loop microstructure design method can be implemented for the unconstrained maximisation of properties such as Specific Surface Area (SSA) and relative diffusivity. Previous approaches have suggested ways of performing a maximisation of certain properties, such as SSA, through a reduction of the particle size. However, this work shows that maximisation of the SSA can be achieved through a rearrangement of the phases while keeping the mean particle size constant. 
Moreover, this work introduces a methodology that defines a constrained function for optimising specific microstructural properties. This method allows for simultaneous optimisation of correlated properties, such as SSA and relative diffusivity. This is crucial for real-world applications where balancing competing requirements is necessary. This was implemented by defining an objective function that accounts for the trade-off between both properties. Similarly, it is possible to maximise these properties while constraining the value of the phase volume fractions to be constant. This approach is key for future works in which the optimisation of microstructural properties is correlated with the experimental synthesis methods. Although the latent space does not directly correspond to physical parameters, linking it to observed microstructural and transport properties could provide a powerful shortcut for identifying optimal configurations, reducing the need for costly experimental data.

The impact of these optimisations should be quantified through full electrochemical simulations over the pristine and optimised microstructures; however, this falls beyond the purpose of this work and is considered relevant to future work. From the generation-optimisation process, it is important to point out that a universally optimum microstructure does not exist since it is closely related to the electrode's purpose. The trade-off between certain properties must be taken into account when defining which property would have a preferred weight over the other during the optimisation. 

By integrating GANs, Gaussian Process, and Bayesian optimization into a deep kernel Bayesian optimization framework, our approach ensures that the designed microstructures are tailored to meet specific application requirements. This adaptability makes the proposed methodology versatile for improving the microstructure of various multiphase electrodes. The ability to design enhanced electrodes is fundamental for advancing energy storage systems as the demand for high-performance energy solutions continues to rise.

\section*{Acknowledgements}

Andrea Gayon-Lombardo acknowledges CONACYT-SENER for financial support. The authors also thank Prof. Stephen Neethling and Dr. Samuel Cooper for their valuable comments.

%Bibliography
\bibliographystyle{unsrt}  
%\bibliography{references}  

\end{document}